\documentclass[prb,reprint, superscriptaddress]{revtex4-2} 

\usepackage{amsmath}  
\usepackage{amsfonts} 
\usepackage{graphicx} 
\setcitestyle{super}

\renewcommand{\Re}{\operatorname{Re}}				
\renewcommand{\Im}{\operatorname{Im}}					

\begin{document}

\title{Frequency-dependent capacitors using paper} 
\author{Bart H. McGuyer}
\altaffiliation{Present address: Facebook, Inc., 1 Hacker Way, Menlo Park, CA 94025, USA}
\affiliation{Department of Physics, Columbia University, New York, NY 10027}

\date{\today}

\begin{abstract}
Measurements of capacitors made with paper sheets reveal a significant decrease of capacitance with increasing frequency from 10 to 100,000 Hz, 
offering a simple demonstration of complex dielectric phenomena using common equipment. 
\end{abstract}

\maketitle

Paper for printing and copying is a convenient material for exploring dielectrics and capacitors.\cite{prevwork1}  
In particular, its shape and uniform thickness is well suited for building parallel-plate capacitors whose capacitance varies with the number of sheets between their plates. 
Paper is a complex 
anisotropic 
dielectric, though, with a variability and environmental sensitivity that can be troublesome for quantitative measurements.\cite{prevwork2}  
However, as Fig.~1 shows, paper often has 
an appreciable dependence on frequency that presents an opportunity to witness complex dielectric phenomena. 

The capacitance $C$ of an ideal parallel-plate capacitor made from two identical plates is 
\begin{align} 
C(f) = \Re \left[ \epsilon_r(f) \right] \, \epsilon_0 \, A / d, 
\end{align} 
where $\epsilon_0$ is the vacuum permittivity, $A$ is the plate area, and $d$ is the thickness of a dielectric between the plates. 
For polarizable dielectrics, the capacitance is enhanced by the real part of the permittivity $\epsilon_r$ (or dielectric constant), which is within roughly 1.3--4.0 for paper.\cite{prevwork2,PaperHandbook,simula:1999,PaperElectronics}  
The closely related imaginary part of $\epsilon_r$ produces loss. 
If a sinusoidal voltage with frequency $f$ is applied, then 
a frequency dependence in the permittivity will create a frequency dependence in the capacitance. 

Many polarizable sources inside paper contribute 
including ions that move and dipoles that rotate, for example, in cellulose and absorbed water.\cite{PaperHandbook,simula:1999,PaperElectronics}  
Their responses are not instantaneous, however, and as the frequency $f$ increases, slow sources contribute less and less. 
For paper, the slowest sources are ionic (space charge) polarizations and their relaxations dominate at low frequency.\cite{PaperHandbook} 
Overlapping of multiple sources blurs distinct features,\cite{simula:1999} leading $C(f)$ to gradually decrease with frequency as shown in Fig.~1(a) (c.f., Fig.~3 of Ref.~4). 
This relaxation also produces loss as shown in Fig.~1(b) (c.f., Fig.~4 of Ref.~4). 
Sharper features occur in other dielectrics or at higher frequencies.

\begin{figure}[b]
\centering
\includegraphics{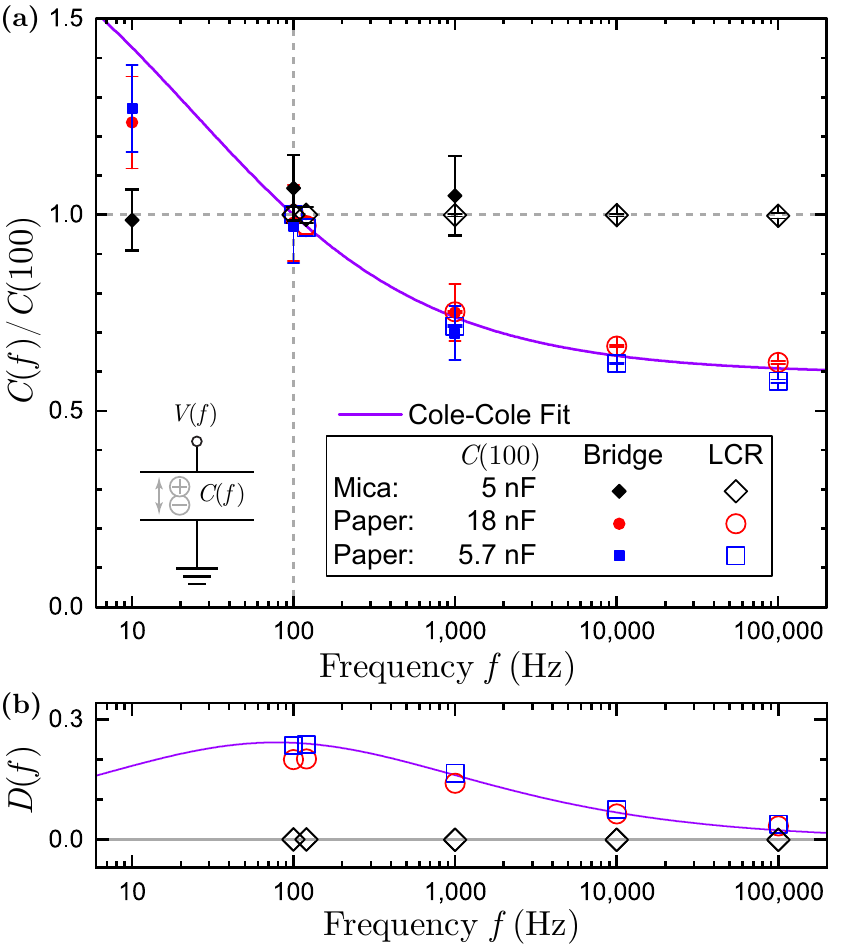}
\caption{
(a) Measured capacitance $C(f)$ versus frequency $f$ for two paper capacitors and a nearly frequency-independent silver mica capacitor. 
Measurements used an impedance bridge or an LCR meter. 
A fitted curve highlights the appreciable variation of the paper capacitors. 
(b) Dissipation factor $D(f)$ from LCR-meter measurements. 
A curve shows the predicted $D$ from the fit of the capacitance data. 
}
\label{schematic}
\end{figure}

To compare the frequency dependence of different capacitors, the data for each capacitor in Fig.~1(a) are normalized using the value at 100 Hz. 
The data for two paper capacitors show a similar variation with frequency, while a control capacitor shows no variation to within measurement uncertainty. 
The loss values in Fig.~1(b) are measurements of a dissipation factor (or loss tangent)  
$D(f) \approx - \Im \left[ \epsilon_r(f) \right] / \Re \left[ \epsilon_r(f) \right] $ for paper over this frequency range.\cite{PaperHandbook,simula:1999,Keysight} 
The data for the paper capacitors again show a similar variation, while the data for the control shows negligible loss.

To highlight the capacitance variation, the data for both paper capacitors were 
fit with the function $y(f) = C(f)/C(100~\text{Hz})$ 
using the Cole-Cole empirical model\cite{cole,review,DA} 
\begin{align}
\epsilon_r(f) &= \epsilon_0 + \frac{ \epsilon_1 }{ 1 + (i f/f_0)^{\gamma} } \\ 
	&= \epsilon_0 + \epsilon_1 \frac{ 1 + [ \cos (\gamma \pi/2) - i \sin (\gamma \pi/2) ] (f / f_0)^{\gamma} }{ 1 + 2 \cos(\gamma \pi/2) (f / f_0)^{\gamma} + (f/f_0)^{2 \gamma} } 
\end{align} 
with $1 \geq \gamma > 0$. 
The second line follows from substituting $e^{i \pi/2}$ for $i$. 
This model captures the broadened relaxation of many solid and liquid dielectrics. 
For $\gamma = 1$, it is equivalent to the Debye model\cite{cole,review,DA,Keysight} for ideal dipolar relaxation. 
While there seems to be no standard model for paper, the fitted curve summarizes the data well. 
However, there was not enough low-frequency data to fully constrain the model. 
Least-squares fitting including uncertainty gave 
$y(\infty) = 0.59 \pm 0.02$, 
$\gamma = 0.50 \pm 0.13$, and 
$f_0 = 24 \pm 94$ Hz. 
The fit predicts a dissipation factor that matches the data well, as shown in Fig.~1(b).

\begin{figure}[t]
\centering
\includegraphics{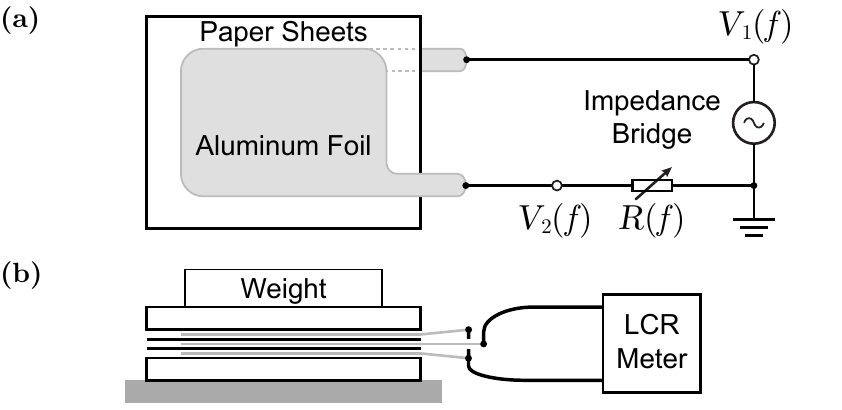}
\caption{
(a) Setup using an impedance bridge with a sketch of a paper capacitor shown from above. 
The bridge was made of a variable resistor $R(f)$ and a function generator producing $V_1(f)$.
(b) Setup using an LCR meter with a sketch of a paper capacitor shown from the side. 
Three plates were used to increase capacitance and reduce fringe fields. 
}
\label{schematic}
\end{figure}

The measurements in Fig.~1 used two paper capacitors made from compressing three aluminum-foil plates inside a standard 500-sheet ream of 30\% recycled copy paper, as sketched in Fig.~2. 
For one capacitor, the neighboring plates were separated by 1 sheet (5.7 nF data), and for the other, by 4 sheets (18 nF data). 
The capacitance is sensitive to plate alignment, fringe fields, ream compression, and air gaps,\cite{prevwork2} which is why the ratio of $C(100)$ values in Fig.~1(a) slightly differs from the ratio of sheets. 

Measurements used 
an impedance bridge as shown in Fig.~2(a) for low frequencies (10--1,000 Hz) and 
an LCR meter (DER EE DE-5000\cite{DER}) as shown in Fig.~2(b) for high frequencies (100--100,000 Hz).  
Either approach alone captured frequency dependence. 
Note that multimeters can often measure capacitance but they tend not to have a well-defined frequency. 
Dissipation factors are typically reported by LCR meters, but an improved, phase-sensitive bridge\cite{VectorBridge} would be needed to measure loss separately from capacitance.

The series circuit of the impedance bridge formed by the capacitor and a 1 M$\Omega$ variable resistor functioned as a voltage divider. 
Using complex values for the voltages and the impedance for an ideal capacitor,\cite{VectorBridge} the capacitance can be calculated from the voltage-dividing ratio $V_2/V_1$ and the resistance $R$ as 
\begin{align} 
C(f) = 1 \big/ \left[ {2 \pi f R(f) \, \sqrt{ {| {V_1(f)}/{V_2(f)}|^{2}} - 1} } \right]. 
\end{align} 
In practice, for each value of $f$ a function generator applied a sinusoidal voltage $V_1(f)$, the resistor was adjusted so that $|V_2(f)/V_1(f)| \approx 1/2$, 
and then an oscilloscope measured the amplitudes  $|V_1(f)|$ and $|V_2(f)|$ 
and a multimeter measured the adjusted resistance $R(f)$. 
Note that a ``true-RMS'' multimeter could replace the oscilloscope. 
The uncertainty $\delta C$ for each value of $f$ follows from a standard propagation of the uncertainties\cite{Bevington} for each parameter, giving 
\begin{align} 
\delta C \approx C \sqrt{ ({\delta R}/{R})^2 + \frac{  |\delta V_1/V_1|^2 + |\delta V_2/V_2|^2 }{ (1-|V_2/V_1|^2)^2  }} 
\end{align}  
assuming negligible frequency uncertainty.

The dielectric properties of materials are fascinating.\cite{review,Keysight}  
Other types of paper or different sheet materials could be explored with this approach. 
I observed similar results with most everyday paper products, from cardboard to a textbook and even a silicone placemat.\cite{silicone}
Commercial paper film capacitors could be used, though their frequency variation is typically only a few \%.\cite{CommercialPaperCap} 
Alternatively, a phase-sensitive bridge\cite{VectorBridge} could be used to separate loss 
from capacitance and reach lower frequencies.\cite{simula:1999} 
Frequency-dependent capacitors like those in Fig.~1 can be modeled as resistor-capacitor networks. 
One prediction of such models is dielectric absorption,\cite{DA} an effect leading some capacitors to recharge over time,   
which is why capacitors must be left shorted after discharging to guarantee safe handling.

I'm grateful to 
Mickey McDonald 
for helpful feedback and encouraging this work, 
and to two anonymous referees for valuable suggestions that improved this work.

\end{document}